# OE-CAM: A Hybrid Opto-Electronic Content Addressable Memory

**Yousra Alkabani, Mario Miscuglio, Volker J. Sorger, Tarek El-Ghazawi**

*Electrical & Computer Engineering Dept.*
*George Washington University, Washington DC*

**Abstract:** A content addressable memory (CAM) is a type of memory that implements a parallel search engine at its core. A CAM takes as an input a value and outputs the address where this value is stored in case of a match. CAMs are used in a wide range of applications including networking, cashing, neuromorphic associative memories, multimedia, and data analytics. Here, we introduce a novel opto-electronic CAM (OE-CAM) utilizing the integrated silicon photonic platform. In our approach, we explore the performance of an experimental OE-CAM and verify the efficiency of the device at 25Gbit/s while maintaining the bit integrity under noise conditions. We show that OE-CAM enables a) two orders of magnitude faster search functionality resulting in b) a five orders of magnitude lower power-delay-product compared to CAMs implementations based on other emerging technologies. This remarkable performance potential is achieved by utilizing i) a high parallelism of wavelength-division-multiplexing in the optical domain, combined with ii) 10's of GHz-fast opto-electronic components, packaged in iii) integrated photonics for 10-100's ps-short communication delays. We further verify the upper optical input power limit of this OE-CAM to be given by parasitic nonlinearities inside the silicon waveguides, and the minimal detectable optical power at the back-end photoreceiver's responsivity given channel noise. Such energy-efficient and short-delay OE-CAMs could become a key component of functional photonic-augmented ASICS, co-processors, or smart sensors.

## 1. Introduction

A content addressable memory (CAM) is a memory with a search engine at its core used to find values stored within the CAM; upon finding a match it responds with the address where this value is stored. This search operation is usually executed within one clock-cycle because the CAM search engine is capable of searching all locations in parallel. Thus, CAMs are especially useful in different applications that demand short search-return and pattern matching. This includes networking and high-speed routing [1], [2], neuromorphic associative memories [3], multimedia and data compression [4], in addition to cashing and data analytics [5], [6].

In a generic CAM a search word is applied at the input of the search engine and the search engine outputs a match line per word that signifies whether the search word is found at this location or not (Fig. 1). The output from the search engine is sent to a priority encoder that generates the address at which the search word is found (or the first address in case of multiple matches). The search engine is composed of CAM cells that are capable of storing data and performing the match operation. A CAM cell is similar to a regular memory cell with the addition of extra logic to perform the comparisons between the stored value and the value applied at the search lines. CAMs can be classified as binary CAMs or ternary CAMs; a binary CAM cell stores either a 0 or a 1, whereas a ternary CAM cell can store *0*, *1*, or a *don't-care*. In this work, we focus on binary CAMs.

As a type of memory, CAM cells are manufactured using SRAM [7]–[12] or DRAM technology [13], [14]. While DRAM based CAMs provide larger capacities, SRAM based CAMs have seen higher adoption as they provide faster search speed, which is a priority for all applications using CAMs. However, both SRAM and DRAM technologies are facing issues due to technology scaling such as i) response-time delays mainly arising from capacitive RC-delay in electronic circuits, and ii) wasteful power dissipation. This prompts exploring the use of novel technology options to design and implement CAMs. For instance, multiple CAM designs were introduced using emerging non-volatile memory such as Phase Change Memories (PCM) [15]–[17], Spin-Transfer Torque Magneto-resistive Random Access Memories (STT-MRAMs) [18]–[20], and memristors [21]–[24], to name a few (Table I). These novel technology options provide the promise of highly scalable and energy-efficient memory functionality, while maintaining acceptable read and write speeds.

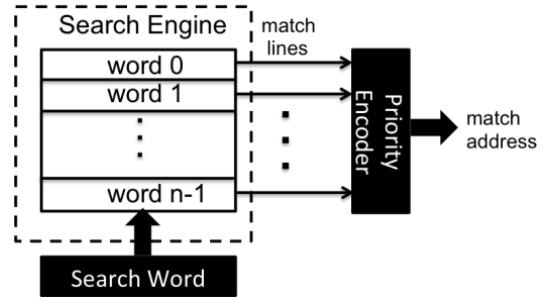

Fig. 1. Block diagram of a generic CAM.

TABLE I
PERFORMANCE OF EMERGING NON-VOLATILE MEMORY TECHNOLOGIES BASED CAMS [22].

|  | SRAM | DRAM | PCM | MRAM | RRAM |
|---|---|---|---|---|---|
| read(ns) | < 0.3 | < 1 | < 60 | < 20 | < 50 |
| write(ns) | < 0.3 | < 0.5 | 60 | 20 | < 250 |

A CAM read and write delay comparison of representative technology options including SRAM, DRAM, PCM, MRAM, and resistive memories (RRAM) shows almost 2-orders of magnitude variability (Table I); in detail, SRAM is (on average) 10x faster compared to all other CAM options, while writes of emerging memory technologies are typically slower and more power-hungry. Note, that in most CAM applications, writes are less frequent than searches, thus, the read speed is of higher importance than the write speed. In addition to speed, other CAM performance metrics include energy efficiency and the chip-footprint (component integration) density. With high search speed in mind, integrated photonics can be a compelling technology option to implement a CAM, as we discuss and show in this work. Exemplary, a fully optical CAM that can operate at 10Gb/s was recently introduced [25], [26], whose design mimics an optical equivalent of a SRAM CAM cell. However, such early attempts of brute-force mapping optical components onto SRAM functionality may not enable the most efficient nor high-performance CAMs. As it stands, the demand for an efficient and low-latency CAM that exploits trusted silicon foundry processes is yet outstanding, and the main rational for our work presented here. Here we introduce a novel opto-electronic CAM design (OE-CAM) that synergistically maps function to technology for the first time. That is, we utilize electronics for the low-speed writes, while performing the reads optically at high-speed. This setting enables satisfying the need of a CAM design to perform both high-speed searches, while allowing for a footprint compact optical CAM cell requiring a low power budget.

While integrated photonics has generally a higher chip footprint per unit functionality, the recently instigated photonic-integrated circuit roadmap anticipates improvement similar to the ITRS, thus technology-scaling and economy-of-scale are expected to enable significant improvements in years to come. The main rationale behind this OE-CAM design presented here is, that a) optics allow for massive data-processing parallelism due to wavelength-division-multiplexing (WDM), and
b) the system's read-response is simply given by the time-of-flight of a photon through the OE-CAM structure; hence, given the compactness of integrated photonics, the read delay is just picoseconds (ps)-short, signifying a 10-100x improvement over electronic-only CAM counterparts.

Figure 2 presents an overview of the proposed OE-CAM; the core of the OE-CAM is an optical search engine that operates at high-speed, where the only delays are given by the opto-electronic device response times (10's GHz-fast) during write operation, and ps-short read delays from the photons time-of-flight. In detail, the search values are stored within the optical search engine by electrically reconfiguring the photonic components (namely microring resonators, MRR, see Sections 3 and 4). This is done by placing the to-be-written word (in binary format) on the

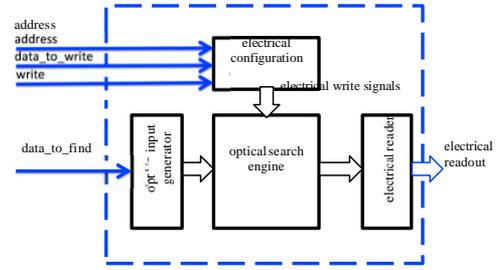

Fig. 2. Overview of the proposed opto-electronic CAM (OE-CAM).

*data to write* lines, placing the location to store the word on the *address* line, then raising the *write* signal. The search word is input electrically to the system using the *data to find* line, which is then generated optically at the input of the optical search engine using laser light. The output of the optical search engine represents the match lines for all the locations, which are converted into electrical signals using photodetectors and whose output is fed to an electrical priority encoder to generate the match address. Incidentally, a practical advantage of this OE-CAM design is its ability to become seamlessly integrated in a standard electrical systems (i.e. plug-in module). However, one could integrate and use OE-CAM also in a photonic integrated circuit architecture by replacing the back-end components (photodetector and electrical decoder) with an optical waveguide-based encoder [27]. In this paper, we focus on the design and evaluation of the optical search engine of OE-CAM.

The remainder of the paper is organized as follows; Section 2, discusses related work. The OE-CAM cell and system design of the OE-CAM search engine are presented in Sections 3 and 4. The new design is evaluated in Section 5. Section 6 concludes the work.

## 2. Related Work

CAMs can be classified as binary and ternary CAMs; a cell of a binary CAM can store either a 0 or a 1, whereas a cell of a ternary CAM can store *0*, *1*, or *don't-care*. In this work, we focus on binary CAMs. Traditionally CAM cells are implemented using regular SRAM cells in addition to matching logic. Two main topologies of SRAM based CAMs are used: NAND and NOR [7]. While the NAND implementation consumes less power than the NOR implementation, its delay increases quadratically with the word size due to capacitive delay limitations (Table I) [7].

A hybrid design combining the advantages of both the NOR and NAND implementations was proposed by Chang and Liao [8]. Other architectural techniques were proposed to reduce the power consumption and improve the throughput of SRAM based CAM cells using pre-computations and overlapped search mechanisms [9]–[12]. Moreover, high-density CAMs with DRAM based cells were proposed, but these CAMs are slower than SRAM CAMs and are thus less attractive [13], [14].

As technology scales, SRAM technology suffers from increased energy consumption due to high leakage currents. Thus, it is essential to study different implementations of CAMs using technological alternatives. Multiple CAM implementations using non-volatile memory technologies were recently proposed, including PCM [15]–[17], STT-MRAMs [18]–[20], and memristors [21]–[24]. While those options promise high scalability, they are all at least one order of magnitude slower compared to SRAM-based CAMs and require rather large write energies [22].

In contrast, here we deploy a promising technology harnessing a variety of benefits found in integrated photonics. Mourgias-Alexandris et al. recently introduced an all-optical CAM that supports read and write operations at up to 10Gbps [25], [26] and a T-CAM based on a similar design was also introduced [28], [29]. Their implementation uses optical components to replicate a traditional SRAM CAM cell and it is based on an optical flip-flop (FF) and an optical XOR gate. At each cell, the FF, responsible for storing the bit, requires two coupled Semiconductor Optical Amplifier-Mach Zehnder interferometer (SOA-MZI) switches in a master-slave configuration that are interconnected through a 5mm-long waveguide. In addition, the output is connected to the lower branch of an MZI which constitutes the XOR gate, comprising two SOA. While being an intriguing concept, in its current implementation, the hardware overhead is rather high. In contrast, our OE-CAM cells are smaller in size than this implementation and do not require any internal light sources. In fact, our OE-CAM cell is composed of one single ring resonator which makes our cell an order of magnitude more compact than the previous optical implementation. Moreover, OE-CAM is fully compatible with the CMOS technology [30], and if needed can be easily integrated within fully electrical systems. Other possible modulator options that can be explored as an alternative to micro ring resonators are Mach-Zehnder modulators [31] that can have efficient attojoule implementations [32].

## 3. Cell Structure

An OE-CAM cell is composed of one silicon based micro-ring resonator (MRR), a component that has gained foundry maturity and as such has been widely described in text books and research papers [33], since they provide a fundamentally important function utilized in several applications ranging from spectral filtering in optical links based on Wavelength Division Multiplexing (WDM) [34], [35] to sensors [36]. To summarize their working principle, when a bus waveguide is coupled to a microring resonator, by means of codirectional evanescent coupling, its transmission spectrum shows a dip near the ring's spectral resonances (Fig. 3). Here, the waves in the ring build up a round-trip phase shift that equals an integer times $2\pi$, and thus interfere constructively. The resonance property can be used as a spectral selectivity provided by the MRR element and is a function of the optical losses ($a$) and coupling coefficient ($r$) (waveguide bus-to-ring), which determines the full width half maximum ($FWHM = \frac{(1-ra)\lambda_{res}^2}{\pi n_g L \sqrt{ra}}$) and Free spectral range ($FSR = \lambda^2/(n_g L)$) of the ring spectrum, setting ring quality factor ($Q$ = FWHM/$\lambda_{res}$ =) and finesse (Finesse = FSR/FWHM). These parameters become determinant for avoiding (spectral) signal cross-talk among neighboring OE-CAM cells since multiple wavelengths are used to express input bits (see below for detailed operation principles).

Moreover, the microrings can be electrically or thermally actuated and therefore act as spectral switches similar to analog modulators. The temperature actuation, however, is rather slow (in $\mu$s scale), while the modulation of portion of the refractive index or the absorption of silicon is influenced by the actual concentration of electrons and holes (i.e. plasma dispersion). The most used MRR modulators, at the foundry and industrial level, therefore include a p-i-n junction that is side coupled to a wave guide as described in [30] or a p-n junction [37]. While carrier injection-based MRR modulators and tunable filters feature high carrier concentrations, their response speed is limited due to long carriers lifetimes (in $ns$ scale). In contrast, depletion-based devices are faster (non limited by the free carrier lifetime), but the signal modulation-depth (i.e. extinction ratio) is typically less pronounced, therefore, the overlap between the optical guided mode and the depletion region has to be maximized. Applying a bias voltage $V_0$ to these MRRs, the transmission spectrum (T) of the ring has a resonant frequency $\lambda_0$. When light passes through the coupled waveguide, the component with wavelength $\lambda_0$ is coupled into the ring. By raising the bias voltage to $V_1$, the resonant frequency shifts to $\lambda_1$ due to the change in the refractive index of the ring, due to either carrier injection or depletion. The difference between $V_0$ and $V_1$ controls the difference between $\lambda_0$ and $\lambda_1$ ($\Delta$).

Another major reason to exploit MRRs is the low energy expended in producing each bit of data, which could be particularly low in the order of few fJ·bit$^{-1}$ [38], [39] while electronics interconnects usually require 23 pJ·bit$^{-1}$. Table II reports the state of the art of silicon modulators based on MRRs. As previously mentioned ring-resonators, based on pn-junctions in reverse bias, provide the best performances in terms of extinction ratio, insertion loss, (reconfiguration, i.e. write) speed and energy consumption along with a relatively compact footprint. In the view of the design of the proposed OE-CAM unit cell, the quality factor of the ring and its free spectral range need to be carefully engineered to match the total number of channels (wavelengths) available, i.e. number of bits which compose a word, and the working spectrum in order to avoid inter-channel crosstalk. Fig. 3 shows the details of an OE-CAM cell and how it can be triggered directly as a lumped element.

TABLE II
STATE OF THE ART SILICON MODULATORS BASED ON MICRO-RING RESONATORS, * [39] IS RELATED TO ER MEASURED IN STATIC CONDITION, LINEAR OPERATION REGION OF 6dB ER ARE GUARANTEED BETWEEN ±3.5$V$

| Material | Device type | IL [dB] | Footprint [$\mu m^2$] | Speed [Gbit/s] | Energy [fJ/bit] | Dynamic ER [dB] |
|---|---|---|---|---|---|---|
| Graphene-Ring [40] | MRR - Loss modulation | Ring dep. | 1600 | 30 | 800 | 3 |
| Silicon [41] | Racetrack - carrier depletion | - | 1000 | 12.5 | 44000 | 1 |
| Si- pn junction [30] | MRR - carrier injection | - | 25 | 12.5 | 8 | 9 |
| Si- pn junction [42] | MRR - carrier depletion | - | 100 | 19 | - | - |
| Si [37] interdigitated pn junction | MRR - carrier depletion | 1 | 2500 | 10 | 287 | 4 |
| Si [38] pn junction | MRR - carrier depletion | 2 | 1000 | 11 | 50 | 6.5 |
| Si pn [43] (IMEC) | MRR - carrier depletion | 3 | 100 | 56 | 45 | 4 |
| Si-Thermal [44] | MRR - carrier depletion thermally tunable | 3 | 100 | 40 | 80 | 7 |
| Si Vertical pn junction [45] | MRR - carrier depletion | 1 | 25 | 25 | 4.25 | 6 |
| Plasmonic [39] | Plasmonic-MRR | 2.5 | 100 | 72 | 12 | 10 (static*) |

Another aspect to take into consideration is the repeatability and consistency of the process used for fabricating the microring-based electro-optic modulators (EOM). Therefore, EOMs based on active materials which require unique integration techniques, not compatible with CMOS processes, are overlooked. Nevertheless, using foundry available Si microring modulators based pn junction and carrier depletion, it is possible to achieve at least 3 dB modulation range at high speed ($> 25$GHz), low energy consumption ($< 50fJ/bit$) compact footprint and contained insertion losses. Plasmonic ring modulators are also ignored due to their lower quality factor which would lead to cross talk penalty. Our OE-CAM cell uses electrical writes and optical reads; to write a 0, we apply a bias voltage of $V_0$ and to write a 1 we change the bias voltage to $V_1$. To detect which value is stored in the cell, we send spectrally different input light signals with $\lambda_i$ where $i \in \{0, 1\}$ into the MRR-based photonic circuit. Now, if no light is detected at the output of the waveguide (i.e. the optical power exciting the optical search engine is below the detector's threshold), then the value $i$ is stored in this cell. Thus, a photodetector can be added at the end of the waveguide to detect a matching value. However, as will be seen in Section 4, we only need one photodetector per OE-CAM entry and thus, the photodetector is not considered as part of the cell.

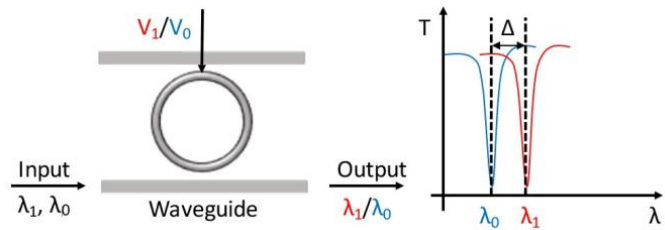

Fig. 3. Cell structure of the CAM. The voltage-controlled read/write operation shifts the MRR resonance which can be cross-references with optical power received by a single back-end photodetector.

# 4. OE-CAM Structure

In this section, we first discuss how OE-CAM cells are combined to construct the optical search engine, followed by showing how multiple OE-CAM search engines can be combined to enable cascadability for constructing larger and parallel-operating search engines.

## 4.1. Construction of OE-CAM optical search engine

To construct an n-bit OE-CAM entry, we need $n$ MRR rings; ring $j$ can be biased to either resonance wavelength $\lambda_{0j}$ or $\lambda_{1j}$. Those resonance wavelengths are spectrally unique (Fig. 4). As previously discussed, the practical word size $n$ is limited by the spectral bandwidth available to tune the ring resonators and the resolution by which we can divide this spectrum.

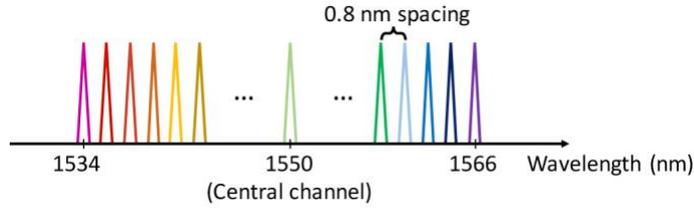

Fig. 4. Resonance wavelength at different locations.

The micro-ring resonators are coupled in series to a waveguide constituting the OE-CAM's entry (row), where each entry is terminated by a photodetector, whose output is connected to a priority encoder (Fig. 5).

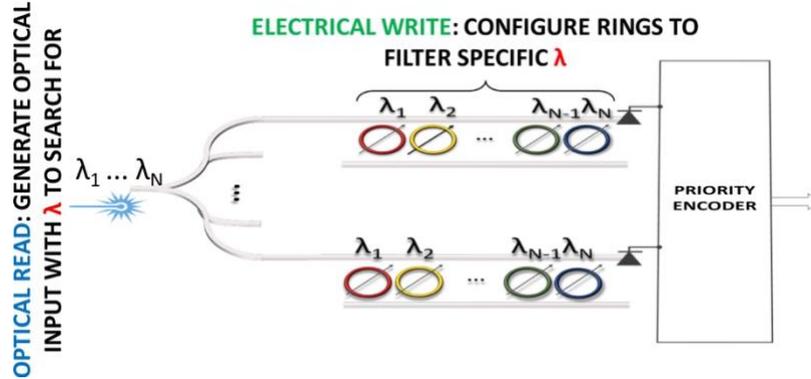

Fig. 5. OE-CAM search engine design.

To store a word in an OE-CAM entry, we apply the proper bias voltage to each cell. For example, if we want to store the word 011.....0, we tune the cells to have resonance frequencies of $\lambda_{00}, \lambda_{11}, \lambda_{12}, ...., \lambda_{0(n-1)}$, respectively. To test whether a word is stored in a specific location or not, we apply light with the corresponding bit-position wavelength components; if no light emerges at the end of that entry's waveguide, then a match is found.

To search for a word, input light composed of those resonance frequencies corresponding to the bits of the word is generated, and send to the MRR filter banks (Fig. 5). The input light will be filtered out only if there is a match. For example to search for the pattern 0101, the input light should include the following wavelengths: $\lambda_{00}, \lambda_{11}, \lambda_{02}$, and $\lambda_{13}$. Although this is a design of a binary CAM, we can search for a don't-care (x) at a specific position by eliminating the light components representing the resonance frequencies of this bit position. Thus, if we want to search for 0X01, the input should be $\lambda_{00}, \lambda_{02}$, and $\lambda_{13}$.

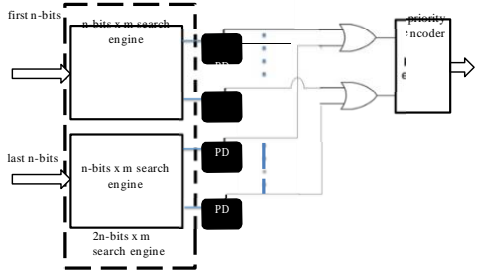
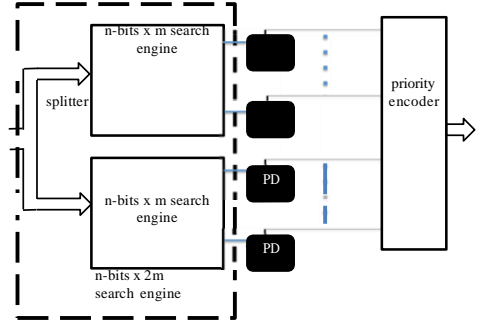

Fig. 6. Combining two optical search engines to construct a wider word search engine.

Fig. 7. Combining two optical search engines to construct a search engine with more entries.

### 4.2. Design Cascadability: construction of larger OE-CAM search engines

Two OE-CAM optical search engines can be combined horizontally to construct an optical search engine to handle larger word sizes as shown in Fig. 6. In this case, the outputs from the photodetector at each entry in the first optical search engine, is combined with the corresponding output in the second optical search engine using an OR-gate. The OR-gate output is then fed to the priority encoder. In this case the read-delay will be increased by a delay equivalent to the delay of a 2-input OR-gate, when compared to an optical search engine that is directly manufactured to be the exact size. Moreover, for search words that are longer than the physical number of MRRs, two optical search engines can be combined vertically to increase the number of entries (Fig. 7). In this case a waveguide splitter (1:N fan-out) is inserted at the input in order to pass the input light to all modules and a large encoder is also needed to accommodate the outputs of the photodetectors in both modules. Here the write and read-delays do not increase (assuming the same number of MRRs per row), thus, the OE-CAM is both wavelength and time-parallelized, and multiple optical search engines can be combined vertically and horizontally to construct optical search engines of any size.

## 5. Evaluation

The read-speed of the proposed design is limited by the time-of-flight of the photon in the photonic integrated circuit and the response-time of the photodetector. Current integrated photodetector technologies can reach up to 50 GHz speed, with a responsivity of at least 0.6 A/W and a noise equivalent power (NEP) of $1\text{pW}/\sqrt{Hz}$ operating in reverse bias (-1V), as shown in table III. The speed of the detector is constrained by the maximum physical length of the device, thus the number of bits of the CAM. Incidentally, similar arguments were made by Tait et al. [46] when discussing MRR-based weight banks for photonic integrated neural network. The length of the OE-CAM is therefore constrained by the pitch of the MRR, where an indicative estimation of the total length $l$ can be obtained via:

$$l = \text{Pitch}_{MRR} \times N_{bit} + l_{Splitter} \quad (1)$$

where $l_{Splitter} = 500\mu\text{m}$ and the $\text{Pitch}_{MRR}$ is 25 $\mu$m. The corresponding time-of-flight of the photon $t$ is:

$$t = 1.6\text{ps} + 0.08\text{ps} \times N_{bit} \quad (2)$$

In order to respect the maximum reading speed of 50 GHz, imposed by the detector, reversing the equation 2, we derive the upper-bound of the number of bits, $N_{bit}$, to be 230 bits, which in our case corresponds to the number of wavelengths. Nevertheless, the number of channels cannot be arbitrary chosen, but it has to respect the total bandwidth available for the WDM modulation and the corresponding spacing between channels (i.e. different wavelengths) is crucial for ensuring signal integrity.

TABLE III
NON-EXHAUSTIVE STATE OF THE ART GRAPHENE, GERMANIUM AND III-V INTEGRATED PHOTODETECTORS ON SILICON.
PLASMONICS, METAL SEMICONDUCTOR METAL, P-I-N JUNCTION IN REVERSE BIAS, AND INP-BASED MODIFIED
UNI-TRAVELING CARRIER (MUTC) ARE CONSIDERED.

| Material | Device type | Responsivity [A/W] | Dark Current [$\mu$A] (Bias (V)) | Speed [GHz] | NEP [pW/$\sqrt{Hz}$] | Footprint [$\mu m^2$] |
|---|---|---|---|---|---|---|
| Germanium [47] | Plasmonic | 0.5 | 20 (-1) | 110 | 1-5 | 0.160 × 10 |
| Graphene [48] | Plasmonic | 0.5 | 4000 (0.4) | 110 | - | 6×1 |
| Germanium [49] | MSM | >0.4 | 4 (5) | 50 | - | 2 × 30 |
| Germanium [50] | MSM | >0.4 | 0.1 (1) | 45 | < 1 | 2 × 10 |
| Germanium [51] | MSM | 0.1 | 86 (1) | 40 | - | 0.7×20 |
| Germanium [52] | p-i-n | 1 | 0.1 (-2 V) | 31 | ≈1 | 4×100 |
| Germanium IMEC [53] | p-i-n | 0.7 | <0.04(-1) | >56 | ≈0.01 | 0.5 × 14 |
| Germanium [54] | p-i-n | 0.8 | 0.003 (-1 V) | 45 | 0.04 | 5.3×8 |
| GaInAsSb [55] | p-i-n | 0.4 | 1.1 (-1 V) | - | 1.5 | 60×60 |
| Germanium [56] | p-i-n | 0.8 | 4 (-1) | 120 | < 3 | 10 × 10 |
| Germanium [57] | MSM | 1 | 300 (5 V) | 25 | - | 10×3.5 |
| InP [58] | MUTC | 0.9 | 0.01 (-5V) | 48 | - | 20×3.5 |

Therefore, our system is designed to exploit Dense-WDM with 40 channels (Fig. 4), leading to a spacing amongst the channels of 0.8 nm, which sets the total number of bits. This is achievable due to strong optical confinement of silicon waveguides using microring resonators with more than 50 nm Free Spectral Range (FSR) with Quality factors ($Q$) close to $10^4$, which allow as many as 50 channels [49], [59]. Assuming approximately 0.8 nm channel spacing, the resonance bandwidth can be broadened up to 0.4 nm, while maintaining an estimated crosstalk level of -10 dB. [49] In this reverse engineering approach, we use the NEP of the detector as reference for estimating the required laser input power and thus the total number of entries (words). Selecting a conservative NEP of 1pW/ $Hz^{1/2}$, we indeed find a detectable (SNR=1) threshold at 50 GHz of 223 nW for each individual channel. Assuming the total number of entries (K), the total power needed to be provided to the OE-CAM can then be computed as:

$$P = K \times P_{\text{Channel}} \times N \quad (3)$$

For avoiding all-optical nonlinearities in the Silicon waveguide platform (e.g. two photon absorption), we consider a conservative 10mW as the maximum optical power running through the system, thus obtaining the total number of entries, $K$ to be 1121. To summarize, an OE-CAM performing reads at 50 GHz could support up to 40 bits using D-WDM and about 1100 entries, ensuring signal integrity maintaining a cross talk level below -10 dB, is realistic requiring a chip footprint of about 28 mm$^2$.

$$\text{Area} = N \times K \times \text{Area}_{\text{MRR}} \quad (4)$$

Regarding the write-speed, diverse technologies and implementations display different performances and modulation efficiency. Even though a detailed discussion is out of the scope of this work, here, we provide a reasonable working range; the write-speed limiting factor in our design is given by the speed at which the refractive index variation occurs in the MRR-based modulators, because of the injection/depletion of carriers, which is used to tune the refractive index. This mechanism is not limited by the recombination times of the generated carriers, but rather by the capacitance of the junction hampered by capacitive effects, that can be minimized at the device engineering stage. It is worth mentioning that capacitive effects can be also used as retention mechanism at the expense of writing speed. Current silicon-based MRR modulators [42], [60] exhibit a speed up to 30 GHz, with a driving voltage of usually few Volts (1-2V) and an efficiency ($V\pi l$) of few tenths of V·cm. Experimental results that corroborate our estimation are reported in [38], where Silicon-based electro-optic MRRs exhibit a modulation in a working spectrum of 0.1 nm and a speed of 11 GHz and as low as 2dB insertion losses. According to our assumptions on the detection bandwidth and modulation efficiency, Table IV summarizes the design parameters of an OE-CAM primarily based on a series of MRRs using the above discussed values from experimental device demonstrations.

TABLE IV
Characteristics of OE-CAM.

| Parameter | OE-CAM |
|---|---|
| cell Elements | MRR |
| Read time (ns) | 0.02 |
| Write time (ns) | 0.03 |
| Write energy (fJ/bit) | 100-200 |
| Density (kbit/cm$^2$) | 160 |
| Operating Voltage | $\approx-2V$ |

Table V shows a comparison of the search power delay product (PDP) of optimized CAM cells based on CMOS, PCM, Magnetic Tunnel Junctions (MTJ), and memristors reported in [17] and our OE-CAM cell. This product is a figure of merit which symbolizes the energy efficiency of the logic mechanism and it is computed as the product between the switching delay (1/50 GHz) and the power required. Our OE-CAM achieves a PDP of just $4.5aJ$ which is an impressive five orders of magnitude savings in search PDP over other technologies. However, thermal interactions between on-chip heat-sources and ring resonators can limit this number, as the micro-ring resonators are extremely sensitive to temperature-induced changes in refractive index. In order to mitigate this effect two main strategies could be considered: (1) Computer aided resonance locking [61]; (2) Thermal-aware Integration strategies [62]. Even if the latter still enables high-speed modulation it requires a great deal of additional power, therefore we believe that integration strategies aiming to minimize cross-talk and thermal effect, using for instance thermal polymers [63] would give a pathway for a high speed modulation using ring resonators without trading off energy consumption. However, we note that there are several emerging materials such as ITO and Graphene that are explored for efficient modulation [64-78], which include new modulation schemes and heterogeneous integration strategies, slow-light effects such as plasmon cavities.

TABLE V
Comparison of the search Power Delay Product [17].

| CAM | CMOS | PCM | MTJ | Memristors | OE-CAM |
|---|---|---|---|---|---|
| PDP ($fJ$) | 14.129 | 36.430 | 52.367 | 15.448 | 0.004 |

We now focus on the simulation of the architecture and its performance. The software used in this analysis is Lumerical Interconnect. We emulate the read- and write-mechanism for one entry of 8 bit. This is intended to be purely illustrative, since the results could be extended to the previously discussed case of a more populated bandwidth. The emulated architecture is represented in Fig. 8. At the input, we minimize the crosstalk penalty by considering individual sources spaced by 1.6 nm (WDM). The first MRR is modulated by a a pseudo-random bit sequence (PRBS) generator at 25 GHz, which codes a sequence of non-return to zero (NRZ) pulses. The quality-factor of the rings

is considered to be 5000, which is practical at the foundry level, with an FSR of 16 nm, and an

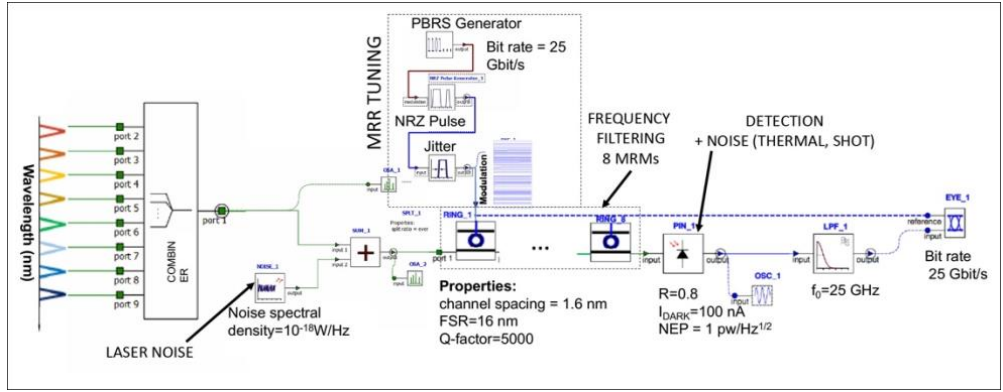

Fig. 8. Hierarchical schematic of a 8-bit OE-CAM. Writing and reading emulation.

operating voltage of 0.5 to -1 V that translates to proportional frequency shifts [53]. Additionally, a noise spectral density ($10^{-18}$ W/Hz) is added to the laser sources. The ring resonator (RING 1) resonant frequency is 193.45 THz (1550.8 nm). The modulation mimics writing in the CAM. (Fig. 9)

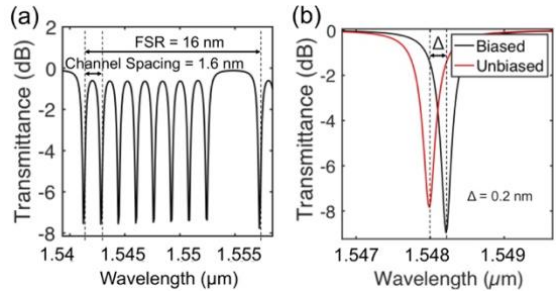

Fig. 9. (a) Transmitted optical power spectrum indicating the ring FSR of 16 nm and channel spacing of 1.6 nm. (b) Transmitted power versus wavelength for biased and unbiased ring modulator.

In a first iteration, the other rings instead behave as passive Bessel-like filters. A jitter source is added to the digital signal which modulates the MRR, to model the short-term variations of the phase. A photodetector converts the analog optical signal into electrical, which is filtered and used as input of bit-error-rate (BER) using as reference the modulation signal (Fig. 10(a)). The detector is considered to be affected by both Thermal (300 K) and shot noise (1 pW/$\sqrt{Hz}$). The eye diagram curves confirm the signal integrity at 25 Gb/s in writing/reading operations. The pattern highlights a width of 35 ps and 1ps time variation of the zero crossing. For just one ring modulated on or off corresponds a Bit error rate (BER) of $4.4 \times 10^{-9}$.

The uncertainty that limits our device scheme is strictly related to the channel bandwidth, in other words, the crosstalk. As reported by Jayatilleka et al [64], based on experimental results in photonic integrated circuit, low inter-channel and intra-channel crosstalk power penalties are demonstrated for MRR filters for data rates up to 20 Gb/s, in this range of proximity [64]. Although, a higher order MRR filter allows reducing concurrent inter-channel and the intra-channel crosstalk, it introduces drawbacks in terms of footprint and power required for the modulation.

Ultimately, we also modulate at the same time 2 MRRs (Fig. 10(b)) and 3 MRRs (Fig. 10(c)), controlled by different pseudo-random bit sequences. Similarly to a PAM modulation, the different levels are still distinguishable, highlighting the robustness of the network. Interestingly, when the resonances of the rings are perfectly aligned to the input wavelengths and concomitantly assuming conservative values of electrical noise at the input, as well as phase jitter, all the bit permutations (such as '01' and '10') fall on the same state 10 b,c. Although, in a real platform implementations, minute fabrication differences can lead to splitting of the permutation states. For low level of noise and cross-talk one can engineer the resonances to discriminate between multiple state arising by the above-mentioned systematic error or by modulation of side wavelengths, with just one adjacent channel, and central wavelengths with 2 adjacent channels for obtaining more information of a partial match.

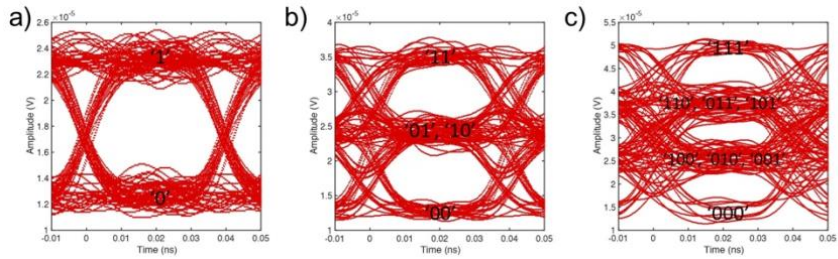

Fig. 10. Output eye diagrams of the EO-CAM for 25 Gbit/s modulation rates of silicon micro-ring modulators (for 1 (a),2 (b) and 3 (c) MRRs) with a Q-factor of 5000 and 1.6nm spacing between channels.

## 6. Conclusion

In this work we have introduced a novel opto-electronic CAM (OE-CAM) with an optical search engine at its core. OE-CAM uses electrical writes to store values in the optical search engine by tuning microring resonators, and achieves high-speed optical search. Our design enables search speeds that are on average 100x faster than CAMs implemented using other emerging technologies. Moreover, it has a search PDP that is five orders of magnitude less than other CAMs. We have verified this performance by simulating an 8-bits OE-CAM using Lumerical Interconnect and verified the efficiency of the device at 25Gbit/s while maintaining the bit integrity under photo detector noise and cross-talk conditions. This high performance opto-electronic search engine using integrated photonics is enabled by a high-degree of parallelism by utilizing wavelength-division-multiplexing, short photon delays inside this relatively compact design, and 10's GHz-fast response times of modern opto-electronic components. Such hybrid search engines are therefore promising candidates for future heterogeneous photonic-electronic integrated circuits.